\documentclass[10pt,letterpaper]{article}

\usepackage[letterpaper,top=0.72in,bottom=0.72in,left=0.82in,right=0.72in]{geometry}
\usepackage[T1]{fontenc}
\usepackage{mathptmx}
\usepackage{microtype}
\usepackage{graphicx}
\usepackage{booktabs}
\usepackage{array}
\usepackage{longtable}
\usepackage{caption}
\usepackage{enumitem}
\usepackage{natbib}
\usepackage[hidelinks]{hyperref}
\usepackage{xcolor}
\usepackage{placeins}
\usepackage{float}

\graphicspath{{figures/}}
\setlist[itemize]{leftmargin=0.52in,itemsep=3pt,topsep=4pt}
\newcommand{\papersection}[1]{%
  \par\vspace{1.1\baselineskip}%
  \begin{center}\normalsize\MakeUppercase{#1}\end{center}%
  \vspace{0.15\baselineskip}}
\newcommand{\uns}{\ensuremath{\Phi_{\mathrm{uns}}}}
\newcommand{\sgn}{\ensuremath{\Phi_{\mathrm{sgn}}}}
\newcommand{\posflux}{\ensuremath{\Phi_{+}}}
\newcommand{\negflux}{\ensuremath{|\Phi_{-}|}}
\newcommand{\kms}{\ensuremath{\mathrm{km\,s^{-1}}}}
\newcommand{\dns}{\ensuremath{\mathrm{DN\,s^{-1}}}}

\begin{document}

\begin{center}
  {\bfseries MULTI-INSTRUMENT ANALYSIS OF NOAA AR 12781: COUPLING SURFACE EVOLUTION OF THE AR\\
  WITH ITS IN-SITU SOLAR WIND SIGNATURES\par}

  \vspace{2.2em}
  Dharanesh Palaniappan\textsuperscript{a,*}, and Khagendra Katuwal\textsuperscript{b}

  \vspace{1.1em}
  \textit{\textsuperscript{a}Madras Christian College, Department of Physics, India}

  \vspace{1.1em}
  \textit{\textsuperscript{b}New Mexico State University, Department of Astronomy, USA}

  \vspace{1.1em}
  \textit{\textsuperscript{a}e-mail: \href{mailto:dharaneshpalaniappan@gmail.com}{dharaneshpalaniappan@gmail.com}}

  \vspace{1.1em}
  \textit{\textsuperscript{b}e-mail: \href{mailto:katuwal@nmsu.edu}{katuwal@nmsu.edu}}
\end{center}

\vspace{1.3em}
\noindent\textbf{Abstract---}Understanding the relationship between photospheric magnetic flux evolution and coronal emission remains an important problem in heliophysics. We investigate the decay of NOAA Active Region 12781 from 7 to 11 November 2020 using observations from the Solar Dynamics Observatory (SDO). Coronal emission was measured from the Atmospheric Imaging Assembly (AIA) 193~\AA{} images, while the photospheric signed and unsigned magnetic flux was derived from Helioseismic and Magnetic Imager (HMI) observations of the Fe~I 6173~\AA{} line. The unsigned magnetic flux and mean 193~\AA{} intensity showed a strong inverse relationship, with Pearson and Spearman correlation coefficients of $r_p=-0.771$ and $r_s=-0.758$, respectively. During the five-day decay phase, the unsigned magnetic flux decreased by 36.54\% from its maximum value, while the coronal emission increased by 21.96\%. We also compared the remote-sensing observations with near-Earth solar-wind and interplanetary magnetic-field measurements from the OMNI database using an estimated propagation delay. The results show that the coronal emission did not vary directly with the photospheric magnetic flux during the decay of AR~12781. The possible association with the solar wind is discussed cautiously because a direct source connection cannot be established from the time-delay estimate alone.

\vspace{1em}

\papersection{Introduction}

The solar magnetic field is highly dynamic and complex due to the differential rotation of the solar interior. This complex magnetic field, along with convection drives a wide range of events such as Coronal Mass Ejections (CMEs), solar flares, coronal holes, and solar prominences which directly affect the space weather. In fact, the magnetic field plays a crucial role in transporting energy from the photosphere to the Corona and even guides the flow of plasma in the solar surface  \citep{Wang1994}. In the solar interior, plasma behaves as a highly conductive, turbulent fluid within the convection zone. This convective motion continuously twists, compresses, shears, and stresses the local magnetic fields via the solar dynamo mechanism. Because these intensely magnetized plasma structures are less dense than their unmagnetized surroundings, magnetic buoyancy forces push the flux tubes upward toward the solar surface. When these magnetic flux tubes cross the photosphere, they forms as localized active regions \citep{BorreroIchimoto2011}. The solar magnetic field is concentrated into localized clusters of magnetic field loops where the magnetic field strength can exceed 1000 G and hundreds of kilometers in diameter \citep{Chitre2003}.These structures appear as dark features known as Sunspots. Sunspots are generally associated with closed magnetic loops, whereas coronal holes are predominantly unipolar magnetic fields \citep{2023AAS...24221404K} and open magnetic-field structures \citep{Harvey1982,2023AAS...24221404K,Katuwal_2026}.

Sunspots can be readily identified in photospheric observations because they appear as dark regions associated with strong, concentrated magnetic fields. In contrast, distinguishing coronal holes from the quiet Sun using photospheric intensity observations alone is considerably more difficult, since both regions generally exhibit weaker magnetic fields than sunspots and active regions. However, their magnetic-field distributions are characteristically different. Coronal holes are typically dominated by magnetic flux of a single polarity, whereas the quiet Sun contains a more balanced mixture of positive and negative magnetic polarities. Therefore, the distribution and degree of imbalance of the line-of-sight magnetic field provide an important means of distinguishing coronal holes from quiet-Sun regions \citep{2023shin.confE..50K,2023AAS...24221404K}. Sunspots and coronal holes are both important sources of space-weather activity, although they influence the heliosphere through different physical processes. Sunspots and coronal holes are both important sources of space-weather activity, although they influence the heliosphere through different physical processes. Coronal holes are major sources of high-speed solar-wind streams. Active regions associated with sunspots can produce solar flares and coronal mass ejections (CMEs) which can release large amounts of plasma and magnetic energy into the heliosphere. In contrast, the quiet Sun exhibits relative stability and is generally unassociated with strong eruptive events or major space-weather disturbances When high-speed solar-wind or CMEs interact with Earth's magnetosphere, they can trigger geomagnetic storms, which affects the satellites, power grids, radio communications, navigation systems, and other technological infrastructure.

Active regions are generally associate with sunspots, and their occurrence varies over the solar maximum and minimum.  A solar cycle lasts approximately 11 years. During solar maximum, the occurrence of sunspots and active regions increases significantly and accompanied by a greater occurrence of energetic phenomena such as solar flares and coronal mass ejections (CMEs), whereas relatively few sunspots appears during solar minimum. The Sun's global magnetic field reverses polarity approximately every 11 years, resulting in a complete magnetic cycle of about 22 years. Similarly, several studies have investigated whether the occurrence, area, location, and magnetic properties of coronal holes exhibit systematic variations with the solar cycle. However, the solar-cycle dependence of coronal holes is more complex than that of sunspots \citep{2026arXiv260823665V}, because their evolution and distribution depend strongly on their latitude, magnetic configuration, and whether they are polar or low-latitude coronal holes.

The sunspots are found to be in the cores of the active regions. Active Regions (ARs) are the major sources for producing solar flares and CMEs \citep{Wang2026,Gough2009}. The study of physical connection between the magnetic flux and the radiative output is essential for understanding how energy is transferred across different layers of the solar atmosphere and for an active region, it highly depends on its specific life cycle. During the initial phase of an active region emergence, the magnetic flux continuously feeds magnetic energy to the system, which yield a highly positive correlation between the magnetic flux and its corresponding EUV coronal radiative output  \citep{Pevtsov2003}.  Once the supply of the new flux loops reduces the active region, enter into its gradual decay phase. During this phase, the organized magnetic fields are no longer driven by the continuous supply of magnetic energy; instead, the existing flux loops are subjected to continuous fragmentation and dispersion driven by the photospheric convective motions. This fragmentation causes the existing magnetic loops to migrate from larger flux concentration to the surroundings. When the opposite polarities of the flux collide or merge, they undergo flux cancellations \citep{Martin1985}. This cancellation neutralizes the net flux and drives the rapid reduction of the unsigned photospheric magnetic flux. During the decay phase, several studies have reported an inverse relationship between the unsigned magnetic flux and the coronal emission, suggesting the magnetic reconnection related activities at the lower altitudes may continuously heat the underlying magnetic loops and drains the magnetic energy reservoirs, thereby maintain high radiative output for a period of time. Once the reconnection and merging of flux starts to fade, it drains the magnetic energy stored under the lower altitudes; the emission gradually reduces to surrounding level.

Although the relationship between magnetic flux and coronal emission is extensively studied during active-region emergence, the evolution of active regions in their non-eruptive decaying phase is less explored. In this work, we present a systematic analysis of the decaying active region AR~12781 using high-resolution data from the Solar Dynamics Observatory. Specifically, we examine temporal variations in coronal brightness using the AIA 193~\AA{} channel and magnetic flux density from HMI. To study the impact of the resulting solar wind, we correlate the remote observations with in-situ plasma data from the OMNI database. Our results suggest a strong inverse relationship between unsigned magnetic flux density and coronal emission during the decay phase of AR~12781.

\papersection{Data}

The remote-sensing observational data used in this study were obtained from the Solar Dynamics Observatory \citep[SDO;][]{Pesnell2012}. Coronal evolution of the active region was analyzed using EUV images from the Atmospheric Imaging Assembly at 193~\AA{} \citep[AIA;][]{Lemen2012}. The 193~\AA{} channel was chosen for active-region analysis because it provides high-contrast observations of coronal plasma at a temperature near 1.5~MK (primarily Fe~\textsc{xii}) \citep{DelZanna2013}. In this channel, active-region structures are highly defined, providing an excellent window to capture coronal dimming, EUV emission, and other radiative characteristics suitable for studying active-region evolution \citep{Vanninathan2015}. The observational window spans 2020 November 7 at 00:00:00~UTC to November 11 at 18:00:00~UTC. To track the intensity variation and structural evolution of the active region across the solar atmosphere, we used AIA full-disk images with a spatial sampling of 0.6~arcsec~pixel$^{-1}$ at a six-hour cadence, resulting in 20 sets of observations. Corresponding magnetogram data were obtained from the Helioseismic and Magnetic Imager to study the evolution of the underlying photospheric magnetic flux \citep[HMI;][]{Schou2012}. Table~\ref{tab:raw} contains the 20 remote-sensing observations. The AIA observations are normalized by their exposure time to convert raw pixel values from data numbers to \dns{}, accounting for variations in exposure time among observations. Table~\ref{tab:normalized} contains the normalized data.

\begin{figure}[!htbp]
  \centering
  \includegraphics[width=0.68\textwidth]{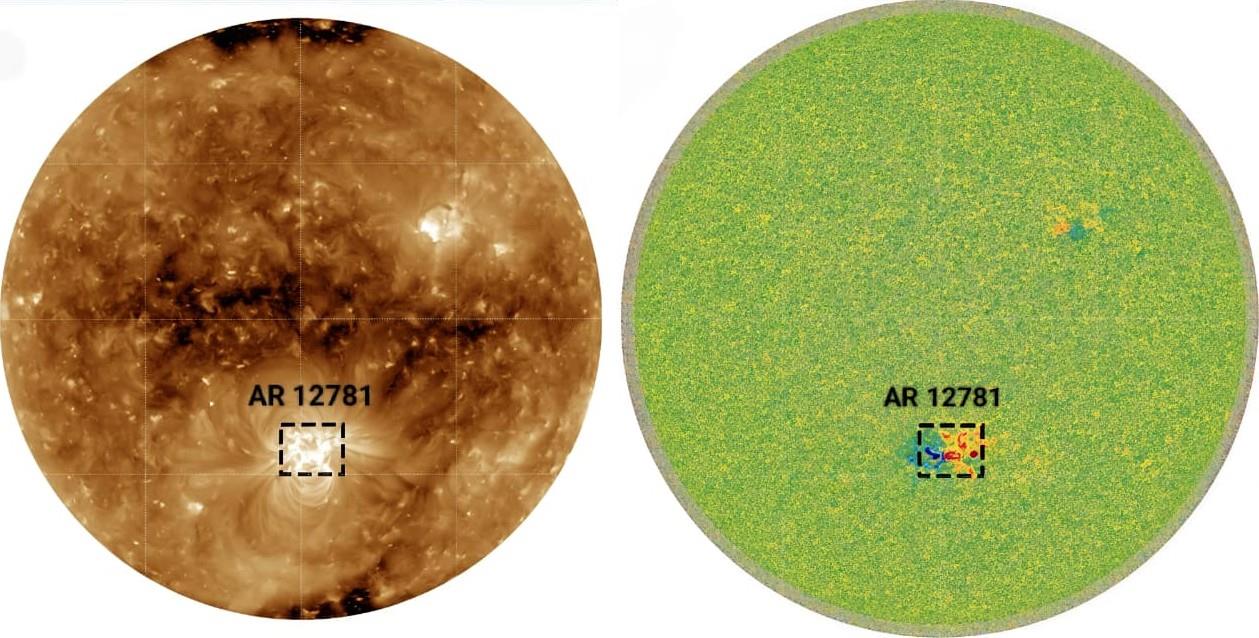}
  \caption{Visualization of AR~12781 in an AIA 193~\AA{} full-disk EUV image (left) with the corresponding HMI 6173~\AA{} photospheric magnetogram (right) on 2020 November 9 at 12:00:00~UTC. The dotted box marks the boundary used to isolate the active region from the quiet-Sun background.}
  \label{fig:ar}
\end{figure}

Additionally, data from the OMNI database were incorporated to analyze the resulting solar wind for in-situ measurements \citep{King2005}. Specifically, we examined the Interplanetary Magnetic Field (IMF) in Geocentric Solar Ecliptic (GSE) coordinates, $B_x$, $B_y$, and $B_z$, along with total magnetic field $|B|$. Bulk plasma parameters such as solar-wind velocity ($V$), proton temperature ($T_p$), proton number density ($N_p$), and flow pressure ($P$) were analyzed to track changes in the plasma. To evaluate the geomagnetic disturbance caused by the resulting solar wind, the symmetric-H (SYM-H) index was obtained. A visualization of AR~12781 is given in Figure~\ref{fig:ar}.

\begin{table}[!htbp]
\centering
\scriptsize
\setlength{\tabcolsep}{5pt}
\renewcommand{\arraystretch}{1.08}
\caption{Temporal variations of AIA intensities and HMI magnetic-flux-density parameters.}
\label{tab:raw}
\begin{tabular}{@{}lccc@{}}
\toprule
Date (UTC) & AIA (DN) & \uns{} (G) & \sgn{} (G) \\
\midrule
2020-11-07 00 & $1192.8780\pm2.6295$ & $119.330\pm0.125$ & $-7.6342\pm0.2215$ \\
2020-11-07 06 & $1282.6028\pm4.2365$ & $122.785\pm0.455$ & $-9.5928\pm0.5570$ \\
2020-11-07 12 & $1276.2614\pm1.5075$ & $122.354\pm0.480$ & $-11.9992\pm0.7010$ \\
2020-11-07 18 & $1350.2518\pm2.8785$ & $126.456\pm0.415$ & $-11.9606\pm0.8910$ \\
2020-11-08 00 & $1297.6786\pm1.7060$ & $124.872\pm0.765$ & $-13.2128\pm1.0750$ \\
2020-11-08 06 & $1269.4956\pm0.9070$ & $123.278\pm0.355$ & $-14.2312\pm0.7400$ \\
2020-11-08 12 & $1316.7814\pm1.7035$ & $120.202\pm0.290$ & $-15.3814\pm0.8190$ \\
2020-11-08 18 & $1323.9620\pm2.4480$ & $120.630\pm0.175$ & $-15.1408\pm0.6580$ \\
2020-11-09 00 & $1298.8450\pm0.9725$ & $117.678\pm0.205$ & $-13.7192\pm0.7700$ \\
2020-11-09 06 & $1326.4870\pm1.9860$ & $116.802\pm0.255$ & $-14.3400\pm0.5120$ \\
2020-11-09 12 & $1305.2908\pm0.1840$ & $114.134\pm0.275$ & $-14.6724\pm0.4635$ \\
2020-11-09 18 & $1334.5346\pm1.5145$ & $112.650\pm0.525$ & $-14.6034\pm0.8090$ \\
2020-11-10 00 & $1379.0792\pm1.6600$ & $108.688\pm0.470$ & $-12.9598\pm1.3230$ \\
2020-11-10 06 & $1439.1702\pm0.6800$ & $106.036\pm0.530$ & $-10.1676\pm1.1750$ \\
2020-11-10 12 & $1442.2664\pm3.7300$ & $100.770\pm0.350$ & $-8.4740\pm0.9050$ \\
2020-11-10 18 & $1449.1100\pm5.7890$ & $96.738\pm0.320$ & $-6.5626\pm0.9500$ \\
2020-11-11 00 & $1436.2754\pm6.5560$ & $92.050\pm0.165$ & $-4.3660\pm0.8300$ \\
2020-11-11 06 & $1435.9842\pm7.6885$ & $88.800\pm0.155$ & $-3.2920\pm0.7900$ \\
2020-11-11 12 & $1398.4532\pm7.9170$ & $84.730\pm0.100$ & $-0.2800\pm0.6400$ \\
2020-11-11 18 & $1406.4828\pm7.0420$ & $80.236\pm0.120$ & $-0.2580\pm0.6900$ \\
\bottomrule
\end{tabular}
\end{table}

\FloatBarrier

\papersection{Methodology}

To study the correlation of photospheric magnetic flux density and coronal radiative emission, we used AIA and HMI data from SDO. Data gathering, coordinate alignment, and spatial reprojection were performed using the SunPy package. To ensure spatial consistency between AIA 193~\AA{} and HMI 6173~\AA{}, spatial reprojection was performed across the two instrument coordinate frames. To isolate AR~12781 from the quiet-Sun background, a localized isolation framework was established. A fixed rectangular boundary with spatial dimensions of $160''\times200''$ was defined on the observation with peak 193~\AA{} intensity and HMI line-of-sight flux density, so the active region remained inside the boundary during the five-day span. This boundary was manually centered over the core of the active region at each of 20 temporal intervals to account for differential rotation. To reduce human error and uncertainty, boundaries were drawn multiple times around the active region to obtain its mean physical parameters. Final data were computed as the arithmetic mean of five replicate measurements. Corresponding error margins for each data point were obtained from the standard deviation of the five measurements. Within the region of interest, total coronal radiative emission and corresponding photospheric magnetic flux density were computed by pixel-intensity summation. While tracking active-region features across the solar disk, geometric foreshortening and line-of-sight projection were mitigated by the low-latitude, near-equatorial trajectory of AR~12781. This process ensured that the primary temporal trends were physically robust.

To study the physical correlation between solar-surface observations and in-situ solar-wind plasma parameters, data were obtained from the OMNI database. Because plasma carried by the solar wind travels at a finite speed through interplanetary space, a temporal propagation delay must be introduced to account for the travel from the solar corona to the instruments. Assuming the average solar-wind velocity, a delay of three days was introduced. The in-situ measurements were examined from 10--16 November 2020, providing sufficient temporal coverage to capture any transient solar streams and interplanetary disturbances caused by the active region.

\begin{table}[!htbp]
\centering
\scriptsize
\setlength{\tabcolsep}{4pt}
\renewcommand{\arraystretch}{1.06}
\caption{Exposure-normalized AIA intensities and HMI magnetic-flux-density data.}
\label{tab:normalized}
\begin{tabular}{@{}lcccc@{}}
\toprule
Date (UTC) & Exp. time (s) & AIA (\dns) & \uns{} (G) & \sgn{} (G) \\
\midrule
2020-11-07 00 & 1.999646 & $596.4390\pm1.3148$ & $119.330\pm0.125$ & $-7.6342\pm0.2215$ \\
2020-11-07 06 & 1.999650 & $641.3014\pm2.1183$ & $122.785\pm0.455$ & $-9.5928\pm0.5570$ \\
2020-11-07 12 & 1.999664 & $638.1307\pm0.7538$ & $122.354\pm0.480$ & $-11.9992\pm0.7010$ \\
2020-11-07 18 & 1.999654 & $675.1259\pm1.4393$ & $126.456\pm0.415$ & $-11.9606\pm0.8910$ \\
2020-11-08 00 & 1.999652 & $648.8393\pm0.8530$ & $124.872\pm0.765$ & $-13.2128\pm1.0750$ \\
2020-11-08 06 & 1.999650 & $634.7478\pm0.4535$ & $123.278\pm0.355$ & $-14.2312\pm0.7400$ \\
2020-11-08 12 & 1.999655 & $658.3907\pm0.8518$ & $120.202\pm0.290$ & $-15.3814\pm0.8190$ \\
2020-11-08 18 & 1.999654 & $661.9810\pm1.2240$ & $120.630\pm0.175$ & $-15.1408\pm0.6580$ \\
2020-11-09 00 & 1.999648 & $649.4225\pm0.4863$ & $117.678\pm0.205$ & $-13.7192\pm0.7700$ \\
2020-11-09 06 & 1.999659 & $663.2435\pm0.9930$ & $116.802\pm0.255$ & $-14.3400\pm0.5120$ \\
2020-11-09 12 & 1.999660 & $652.6454\pm0.0920$ & $114.134\pm0.275$ & $-14.6724\pm0.4635$ \\
2020-11-09 18 & 1.999660 & $667.2673\pm0.7573$ & $112.650\pm0.525$ & $-14.6034\pm0.8090$ \\
2020-11-10 00 & 1.999644 & $689.5396\pm0.8300$ & $108.688\pm0.470$ & $-12.9598\pm1.3230$ \\
2020-11-10 06 & 1.999648 & $719.5851\pm0.3400$ & $106.036\pm0.530$ & $-10.1676\pm1.1750$ \\
2020-11-10 12 & 1.999651 & $721.1332\pm1.8650$ & $100.770\pm0.350$ & $-8.4740\pm0.9050$ \\
2020-11-10 18 & 1.999650 & $724.5550\pm2.8945$ & $96.738\pm0.320$ & $-6.5626\pm0.9500$ \\
2020-11-11 00 & 2.000076 & $718.1377\pm3.2780$ & $92.050\pm0.165$ & $-4.3660\pm0.8300$ \\
2020-11-11 06 & 2.000079 & $717.9921\pm3.8443$ & $88.800\pm0.155$ & $-3.2920\pm0.7900$ \\
2020-11-11 12 & 2.000080 & $699.2266\pm3.9585$ & $84.730\pm0.100$ & $-0.2800\pm0.6400$ \\
2020-11-11 18 & 2.000080 & $703.2414\pm3.5210$ & $80.236\pm0.120$ & $-0.2580\pm0.6900$ \\
\bottomrule
\end{tabular}
\end{table}

\FloatBarrier

\papersection{Dataset and Trends}

The systematic analysis of NOAA AR~12781 during its five-day transit from 2020 November 7 to November 11 provides a valuable dataset capturing a decaying active region. The unsigned line-of-sight magnetic flux density (\uns), signed magnetic flux density (\sgn), and coronal 193~\AA{} emission of AR~12781 were obtained from AIA and HMI as described in the methodology. The total unsigned magnetic flux density \uns{} quantifies the bulk magnetic-field concentration. Net magnetic flux density is the difference between the total positive and total negative flux-density components. The raw parameters, including mean 193~\AA{} emission, \uns, and \sgn, are given in Table~\ref{tab:raw}; exposure-normalized parameters are given in Table~\ref{tab:normalized}.

\begin{figure}[H]
  \centering
  \includegraphics[width=0.68\textwidth]{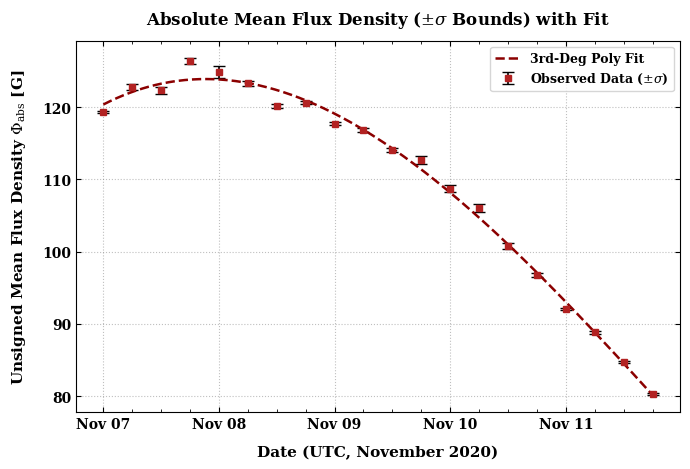}
  \caption{Temporal variation of the unsigned line-of-sight photospheric magnetic flux density (\uns) of NOAA AR~12781 obtained from SDO/HMI magnetograms.}
  \label{fig:unsigned}
\end{figure}

Quantitative evaluation of the SDO parameters reveals that AR~12781 exhibits two distinct evolutionary patterns simultaneously. The \uns{} measurements reach a peak of $126.456\pm0.415$~G on November 7 at 18:00~UTC. Following this peak, the magnetic-field concentration undergoes a steady, monotonic reduction, reaching a minimum of $80.236\pm0.120$~G on November 11 at 18:00~UTC (Figure~\ref{fig:unsigned}). This represents an approximately 36.55\% reduction in \uns{} over the five-day window, confirming that the active region is undergoing its decay phase.

\begin{figure}[H]
  \centering
  \includegraphics[width=0.68\textwidth]{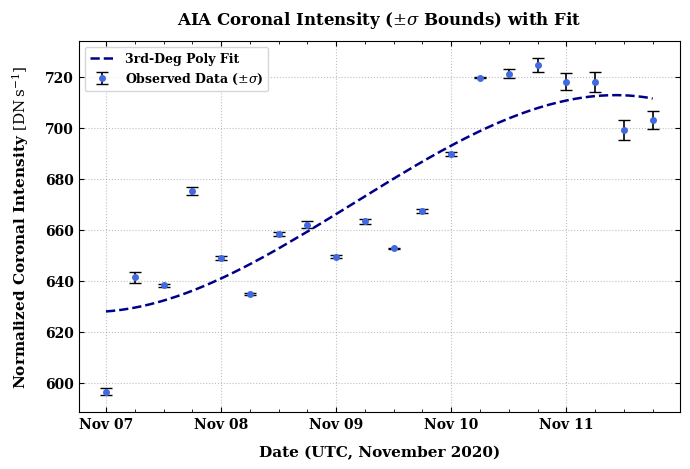}
  \caption{Temporal evolution of the exposure-normalized coronal 193~\AA{} emission intensity associated with NOAA AR~12781 observed in the AIA 193~\AA{} channel.}
  \label{fig:aia}
\end{figure}

In contrast, the coronal radiative output exhibits a strong upward trend. The normalized AIA 193 Å intensity initially begins with a minimum value of $596.5446\pm1.3150$~\dns{} on November 7 at 00:00 and increases to a maximum of $724.6818\pm2.8950$~\dns{} on November 10 at 18:00~UTC, an increment of approximately 21.48\% (Figure~\ref{fig:aia}). These two distinct patterns confirm a nonlinear relationship between the coronal radiative output and the line-of-sight\uns. This distinct nonlinear behavior is a classic signature of a decaying active region. As the magnetic structures fragment and disperse into the surroundings, the stored magnetic energy is converted into thermal energy \citep{Klimchuk2006}, which appears as bright structures in remote observation. This continuous dissipation drains the bulk magnetic footprint while heating the coronal loops to higher temperatures  \citep{Aschwanden2007}. This process maintains the elevated brightness until the magnetic reservoirs are sufficiently depleted, after which both the flux and coronal intensity begin to fade and gradually merge with the background quite-Sun network \citep{Schrijver1997}.

\begin{figure}[H]
  \centering
  \includegraphics[width=0.68\textwidth]{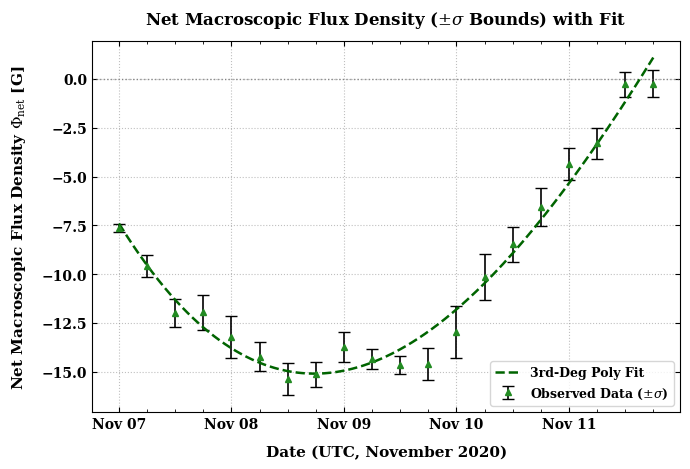}
  \caption{Signed photospheric line-of-sight flux-density (\sgn) asymmetry evolution of NOAA AR~12781.}
  \label{fig:signed}
\end{figure}

In addition to the total unsigned magnetic flux density, the line-of-sight signed flux density \sgn{} provides insight into the active region's polarity asymmetry and structural evolution. During the five days of observation, \sgn{} is dominated by negative polarity. It exhibits a non-monotonic trend (Figure~\ref{fig:signed}) given in Table~\ref{tab:normalized}. Initially, \sgn{} was $-7.6342\pm0.2215$~G on November 7 at 00:00~UTC. It reached a peak negative imbalance of $-15.3814\pm0.8190$~G on November 8 at 12:00~UTC and reduced to $-0.2580\pm0.6900$~G at the end of the observation. To differentiate the physical drivers of the trend, \sgn{} was partitioned into positive and negative flux-density components, compiled in Table~\ref{tab:polarity}.

\begin{table}[!htbp]
\centering
\scriptsize
\setlength{\tabcolsep}{6pt}
\renewcommand{\arraystretch}{1.08}
\caption{Separated positive and negative magnetic-flux components for AR~12781.}
\label{tab:polarity}
\begin{tabular}{@{}lcc@{}}
\toprule
Date (UTC) & Positive flux \posflux{} (G) & Negative flux \negflux{} (G) \\
\midrule
2020-11-07 00 & $55.8479\pm0.1733$ & $63.4821\pm0.1733$ \\
2020-11-07 06 & $56.5961\pm0.5060$ & $66.1889\pm0.5060$ \\
2020-11-07 12 & $55.1774\pm0.5905$ & $67.1766\pm0.5905$ \\
2020-11-07 18 & $57.2477\pm0.6530$ & $69.2083\pm0.6530$ \\
2020-11-08 00 & $55.8296\pm0.9200$ & $69.0424\pm0.9200$ \\
2020-11-08 06 & $54.5234\pm0.5475$ & $68.7546\pm0.5475$ \\
2020-11-08 12 & $52.4103\pm0.5545$ & $67.7917\pm0.5545$ \\
2020-11-08 18 & $52.7446\pm0.4165$ & $67.8854\pm0.4165$ \\
2020-11-09 00 & $51.9794\pm0.4875$ & $65.6986\pm0.4875$ \\
2020-11-09 06 & $51.2310\pm0.3835$ & $65.5710\pm0.3835$ \\
2020-11-09 12 & $49.7308\pm0.3693$ & $64.4032\pm0.3693$ \\
2020-11-09 18 & $49.0233\pm0.6670$ & $63.6267\pm0.6670$ \\
2020-11-10 00 & $47.8641\pm0.8965$ & $60.8239\pm0.8965$ \\
2020-11-10 06 & $47.9342\pm0.8525$ & $58.1018\pm0.8525$ \\
2020-11-10 12 & $46.1480\pm0.6275$ & $54.6220\pm0.6275$ \\
2020-11-10 18 & $45.0877\pm0.6350$ & $51.6503\pm0.6350$ \\
2020-11-11 00 & $43.8420\pm0.4975$ & $48.2080\pm0.4975$ \\
2020-11-11 06 & $42.7540\pm0.4725$ & $46.0460\pm0.4725$ \\
2020-11-11 12 & $42.2250\pm0.3700$ & $42.5050\pm0.3700$ \\
2020-11-11 18 & $39.9890\pm0.4050$ & $40.2470\pm0.4050$ \\
\bottomrule
\end{tabular}
\end{table}

This analysis reveals that both positive and negative flux density decrease over time. Positive flux density reduced from $55.8479\pm0.1733$~G to $39.9890\pm0.4050$~G (a 28.39\% reduction), while negative flux density reduced from $63.4821\pm0.1733$~G to $40.2470\pm0.4050$~G (a 36.60\% reduction). This simultaneous reduction provides evidence for ongoing magnetic-flux cancellation and magnetic dispersion into the background quiet-Sun network \citep{Martin1985}. Photospheric convective motions actively drive opposite magnetic polarities toward the polarity inversion line, causing neutralization and magnetic reconnection in the atmosphere \citep{Chae2004,VanDriel2015}.

\FloatBarrier

\papersection{Correlations}

To quantify the statistical correlation between line-of-sight \uns{} and the corresponding coronal response, we computed the Pearson linear correlation coefficient ($r_p$) and Spearman rank correlation coefficient ($r_s$). This evaluation provided a strong inverse correlation between \uns{} and exposure-normalized coronal 193~\AA{} emission. The Pearson coefficient yields $r_p=-0.771$, supported by a Spearman coefficient of $r_s=-0.758$. To study the evolutionary pattern, both parameters were correlated against time. The temporal correlation of \uns{} yields robust negative coefficients of $r_p=-0.929$ and $r_s=-0.934$, confirming continuous magnetic structural changes. In contrast, the temporal correlation of coronal 193~\AA{} emission yields strong positive coefficients of $r_p=0.863$ and $r_s=0.839$, suggesting steady intensification in the radiative output. Taken together, the statistical analysis supports the interpretation of a decaying active region. The systematic reduction in bulk magnetic flux density and intensification in coronal output are consistent with continuous magnetic-flux cancellation, flux dispersion, and subsequent reconnection of magnetic-flux structures.

\papersection{In-Situ Measurements and Geomagnetic Response}

\begin{figure}[H]
  \centering
  \includegraphics[width=0.64\textwidth]{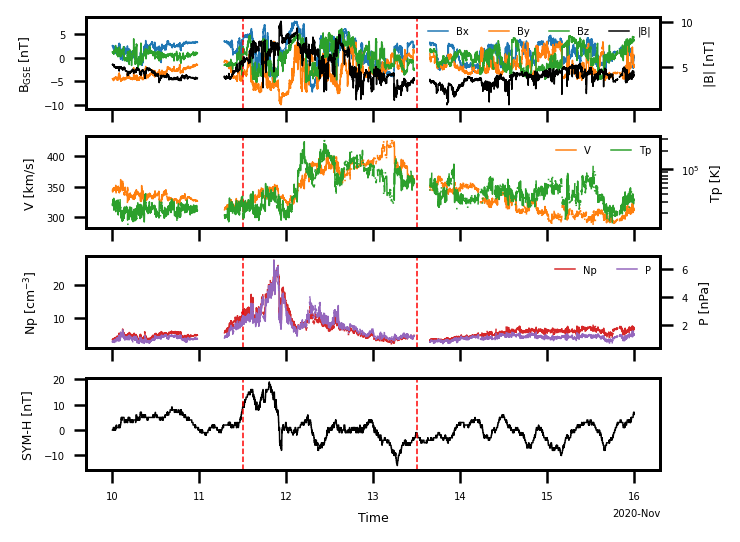}
  \caption{Temporal variation of interplanetary magnetic-field, solar-wind plasma, dynamic-pressure, and SYM-H parameters from the OMNI database.}
  \label{fig:omni}
\end{figure}

To investigate properties of the possible solar-wind signature of decaying AR~12781, in-situ observations from the OMNI database were evaluated. Considering the typical interplanetary propagation delay from the solar surface to instruments at 1~AU, the observation window was shifted to 2020 November 10--16. Within this interval, a distinct pattern is observed between November 11 at 12:00~UTC and November 13 at 12:00~UTC in the interplanetary parameters shown in Figure~\ref{fig:omni}. On November 11, a variation in IMF $B_z$ from approximately $-5$ to $+5$~nT is recorded. Within this interval, the bulk solar-wind velocity averages 362.5~\kms{} and peaks at 425.75~\kms{} on November 13 at 05:03~UTC, indicating a possible stream interaction. The OMNI dataset records a sudden elevation in solar-wind temperature. Correspondingly, proton density peaks near 25~cm$^{-3}$, exceeding typical quiet-solar-wind values. Because dynamic flow pressure is directly related to proton density and the square of bulk velocity, a synchronized sharp spike in dynamic flow pressure is recorded. Such behavior is consistent with enhanced coronal activity in the remote-sensing observations. The elevated temperature and proton density, together with slow bulk speed, are characteristic of a Corotating Interaction Region (CIR)-like structure rather than an eruptive solar phenomenon. Cross-checking against the Interplanetary Coronal Mass Ejection catalog indicates that no CME was recorded during this period. The synchronized spikes in proton density and dynamic flow pressure on November 12 are consistent with compression at a stream interaction region.

The global geomagnetic response is recorded by the SYM-H index. Between November 11 and 12, SYM-H reaches approximately $+20$~nT and subsequently drops to approximately $-15$~nT by November 13. This pattern suggests that the solar wind induced weak magnetospheric compression and a minor geomagnetic disturbance. These variations are consistent with a CIR driven by a weak high-speed solar-wind stream, possibly associated with coronal holes in the vicinity of decaying AR~12781. Because a fixed delay alone does not establish source connectivity, this association remains tentative.

\FloatBarrier

\papersection{Summary and Conclusions}

This study links the evolution of a decaying solar Active region NOAA AR 12781. The high-resolution remote sensing data from the Solar Dynamics Observatory was obtained for the five days transit period of the Active Region across the solar atmosphere and the clear co evolution between the \uns{}  and the coronal radiative heating was tracked. As the active region transited across the solar disk, it underwent magnetic cancellation and dispersion, transitioning from a highly asymmetric, negative polarity dominated region to a near magnetically neutral state. By introducing an offset of three days for the propagation of the solar wind from the sun to the instrument, we were able to correlate the solar surface changes with the in-situ measurements from the OMNI database. The resulting solar wind associated with the active region NOAA AR 12781, enhanced the proton density and elevated temperature that compressed the Earth’s magnetosphere, inducing a minor geomagnetic storm

\begin{itemize}
  \item This research demonstrates strong, statistically valid inverse correlation between the \uns{} and the coronal 193 Å emission during its decay phase. During the five days of its transit, the magnetic flux density dropped by 36.55\%,  while the coronal brightness increased by 21.48\%.
  \item The positive and negative components of magnetic flux density decrease simultaneously, supporting the interpretation that the active region was undergoing magnetic cancellation and dispersion into the background solar atmosphere.
  \item At the beginning of the observation, \sgn{} was dominated by negative polarity; by the end of the window it had dropped to a near-zero value of $-0.2580\pm0.6900$~G.
  \item The OMNI interval contains high-density solar wind with an average velocity of 362.5~\kms{} and a peak velocity of 425.75~\kms{} after applying the approximate propagation delay.
  \item The near-Earth response includes an excursion to approximately $+20$~nT in SYM-H, consistent with weak magnetospheric compression and a minor geomagnetic disturbance. The solar-source association remains tentative.
\end{itemize}

\papersection{References}
\begingroup
\renewcommand{\refname}{}
\vspace{-2.8em}
\setlength{\bibsep}{2pt}
\bibliographystyle{unsrtnat}
\bibliography{references}
\endgroup

\end{document}